# Low-Moment Semiconducting Properties of Quaternary Heusler Alloy CoRuTiSn

Ravinder Kumar[1], Tufan Roy[2], Masafumi Shirai[2,3] and Sachin Gupta[1*]

[1]*Department of Physics, Bennett University, Greater Noida 201310, India*
[2]*Center for Science and Innovation in Spintronics, Core Research Cluster, Tohoku University, Sendai 980-8577, Japan*
[3]*Research Institute of Electrical Communication, Tohoku University, Sendai 980-8577, Japan*

**Abstract**

We investigate structural, magnetic and transport properties of CoRuTiSn equiatomic quaternary Heusler alloy. CoRuTiSn was synthesized by arc-melt technique. The room temperature powder XRD pattern was analyzed, and it was found that CoRuTiSn has a tetragonal crystal structure. Magnetic measurements show non-zero but small hysteresis indicating CoRuTiSn as a soft ferromagnetic with a Curie temperature of ~200 K. The magnetic moment determined from magnetization data is found to be 0.84 $\mu_B/f.u.$ at 4 K, which is close to the value, calculated using first principles calculations. Electrical resistivity decreases with temperatures, indicating semiconducting nature of CoRuTiSn. Hall effect measurements show anomalous behavior, consistent with the ferromagnetic nature of the sample. The low moment ferromagnetic semiconducting nature of CoRuTiSn could make this material promising for semiconducting spintronics.

*Corresponding author email: gsachin55@gmail.com

## 1. Introduction

Significant progress has been made in the field of semiconductor spintronics; however, the exploitation of its complete potential in mainstream technology still encounters considerable challenges.[1–4] One of the key challenges is the efficient injection of spin-polarized current into semiconductors, which is essential for various spintronic applications.[5-6] Among the various strategies explored to address this challenge, spin filter materials (SFMs) have attracted significant attention due to their promising properties.[7-8] Magnetic semiconductors can be highly promising spin filter materials for efficient spin injection into semiconductors due to their ability to selectively filter spin-polarized electrons and their low impedance mismatch.[9-10] Fig. 1 (a and b) shows the schematics of density of states (DoS) for a semiconductor and a magnetic semiconductor. Like semiconductors, magnetic semiconductors have a gap in both spin channels; however, the gaps are unequal, enabling these materials to selectively filter spin-polarized electrons. Various materials such as diluted magnetic semiconductors[11-12], oxide-based magnetic semiconductors[13], spin gapless semiconductor[14–16] and two-dimensional materials[17] have been explored as magnetic semiconductors. Spin gapless semiconductors exhibit unique band structures, where one spin channel has a finite gap, while in the other spin channel, the valence and conduction bands touch at the Fermi level [Fig. (1(c)].[18] These unique band structures can lead to very high spin polarization. As research advances, scientists are striving to develop new magnetic semiconductors, opening new avenues for spintronic devices that could revolutionize future technologies by enabling faster, more efficient, and highly integrated systems.

Heusler alloys offer great potential for investigating magnetic semiconducting properties, owing to their high Curie temperature and large spin polarization. Moreover, their tunable electronic structures enable precise control over magnetic and electronic properties.[19–21] Among these, quaternary Heusler compounds with their complex atomic arrangements and enhanced structural and functional tunability, provide a versatile framework for designing materials with tailored magnetic and semiconducting properties.[22–25] Various quaternary Heusler alloys have been reported theoretically to show magnetic semiconducting properties, however a few have been experimentally realized.[7,21,26–28] In this paper, we synthesize CoRuTiSn quaternary Heusler alloy using arc-melt technique and investigate its structural, magnetic and transport properties along with theoretical calculations. The material exhibits a unique combination of low magnetic moment (~0.84 $\mu_B$/f.u. at 4 K), a narrow bandgap (~60.4 meV), and soft ferromagnetic behavior up to ~200

K. These properties suggest its potential as a SFM, enabling efficient spin-polarized current injection into semiconductors with reduced impedance mismatch due to its semiconducting nature. Our study provides a strong base for further investigations to tune the physical properties for room temperature spintronic applications.

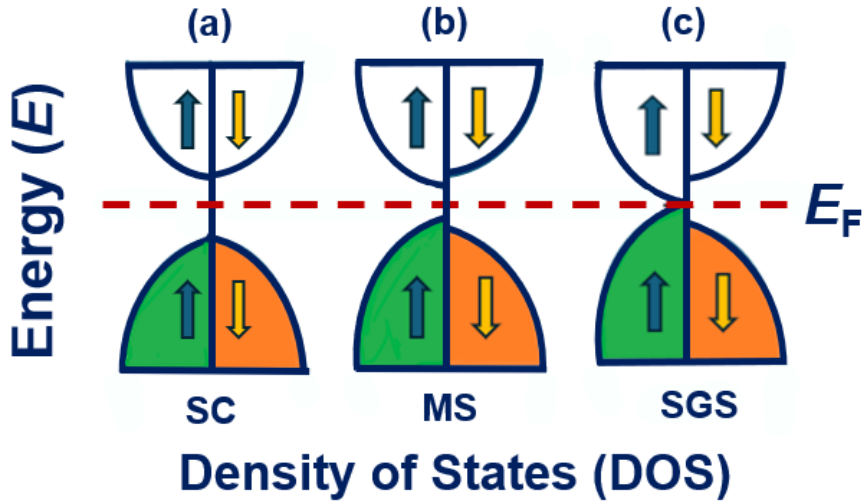

**Fig. 1.** Schematic of energy versus density of states (DOS) for different types of semiconductors; (a) non-magnetic semiconductor (SC), (b) magnetic semiconductor (MS), and (c) spin gapless semiconductor (SGS).

## 2. Experimental and computational methods

A polycrystalline sample of CoRuTiSn was prepared in a high purity argon environment using arc melt technique. Constituent elements Co, Ru, Ti, Sn with atomic purity better than 99.99 %. were taken in a stoichiometric ratio of 1:1:1:1 and melted in the copper hearth. To further improve structural order, the sample was enclosed in a vacuum-sealed quartz tube and annealed at 850 °C for 7 days and then quenched in ice water. To identify the crystal structure of sample, an X-ray diffraction (XRD) pattern was obtained using Cu Kα radiation ($\lambda$ = 1.54 Å) on a Bruker D8 Advance diffractometer at room temperature. A vibrating sample magnetometer (VSM) associated with physical property measurement systems (PPMS) (Cryogenic Limited, UK) was used to measure magnetization of the sample. Electrical transport measurements were performed using a four-probe method by applying 5 mA current in PPMS.

First-principles calculations were performed using projector augmented wave method as implemented in Vienna Ab initio Simulation Package.[29–31] We used generalized-gradient-approximation (GGA) for the exchange correlation energy as parameterized by Perdew, Burke and Ernzerhof.[32] We set plane wave cutoff energy to 520 eV in all the calculations. A k-mesh of 12 ×12 ×12 was used for self-consistent-field (scf) calculations. The effect of swapping disorder in CoRuTiSn was studied using a 128-atoms supercell. The randomness of substitution was considered by adopting special-quasirandom-structures (SQS) as implemented in Alloy Theoretic Automated Toolkit (ATAT) package.[33-34]

## 3. Results and Discussion

### 3.1 Crystal structure

To study the crystal structure of CoRuTiSn quaternary Heusler alloy, X-ray diffraction (XRD) measurement was carried out at room temperature using Cu $K\alpha$ ($\lambda$ = 1.54 Å) radiation in the $2\theta$ range of 20 - 80 degrees. The obtained XRD pattern was analyzed using FullProf suite, which utilizes the least-square refinement between experimental and calculated intensities.[35] Fig. 2 shows powder XRD pattern for CoRuTiSn along with the Rietveld refinement. The bottom plot shows the difference between experimental and calculated XRD patterns. The Rietveld refinement reveals the tetragonal structure with a space group *P42/nnm* (134). It can be noted that the XRD pattern shows a few unwanted peaks, which are predicted to be from small impurity of $Co_3Sn_2$ (less than 8% as per Rietveld refinement). Tetragonal phases were also observed previously in some Heusler alloys such as $Mn_{3-x}Co_xGa$ and $Mn_{3-x}Fe_xGa$.[36-37] Transition of the cubic phase into tetragonal phase may attributed to stress/strain effect along the *c* axes.[38] Lattice parameters obtained from the Rietveld refinement are found to be *a* = *b* = 6.21 Å, and *c* = 6.14 Å. It has been reported that any disorder, defect, stress or strain can lead to change in the magnetic and transport properties of the material.[37-39-40]

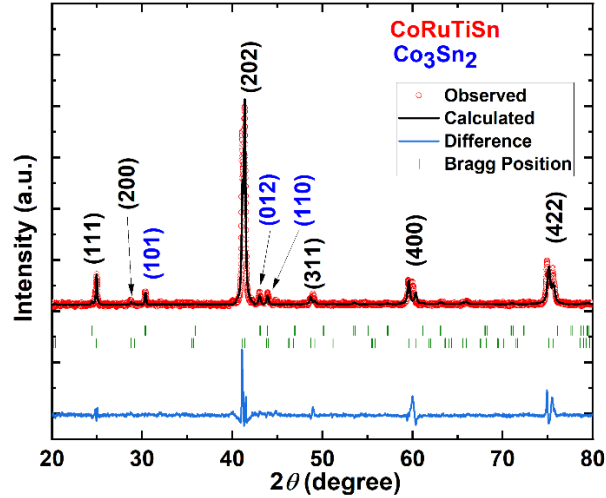

**Fig. 2.** Powder X-ray diffraction pattern with Rietveld refinement for CoRuTiSn taken at room temperature. Impurity peaks of $Co_3Sn_2$ are indexed in blue color.

### 3.2 Magnetic properties

Fig. 3(a) shows the thermo-magnetic (*M-T*) curves obtained in an applied field of 500 Oe for CoRuTiSn under zero-field cooled (ZFC) and field-cooled (FC) configurations. The Curie temperature of the CoRuTiSn (derived from the first-order derivative of *M* vs. *T* plot) was found to be ~ 200 K as indicated by an arrow. Figure 3(a) illustrates that both ZFC and FC curves exhibit a gradual increase in magnetization as the temperature decreases below the Curie temperature ($T_C$), reflecting the alignment of magnetic moments as thermal energy diminishes. The gradual increase in magnetization well below the Curie temperature follows Bloch's $T^{3/2}$ law.[41] Similar to other quaternary Heusler alloys (e.g., CoRuMnGe), CoRuTiSn shows a slight upturn in the *M-T* plot at the lowest temperatures.[42] The upturn in magnetization at low temperatures could be attributed to the following factors: (i) the suppression of spin-wave excitations, which decrease at low temperatures, driving magnetization toward saturation;[43] and/or (ii) establishment of long range order in the $Co_3Sn_2$ impurity phase. Fig. 3(b) shows the field, *H* dependence of magnetization, *M* as a function of temperature, *T*. The *M-H* data was recorded at 4, 100, 200 and 300 K temperatures. It can be observed from Fig. 3(b) that CoRuTiSn shows ferromagnetic behavior up to 200 K reflected by S - shape curve. At 300 K, the sample becomes paramagnetic as indicated by linear magnetic field dependence. The magnetic moment determined at 4 K is found to be ~ 0.84 $\mu_B$ /f.u, which is slightly lower than the theoretical value calculated from the Slater–Pauling rule.[28]

Slater-Pauling rule predicts saturation magnetic moment of 1 $\mu_B/f.u$ in CoRuTiSn. Tetragonal distortion in the crystal structure of the material might account for deviation in magnetic moment from the calculated value.[39] It has been reported that $Co_3Sn_2$ exhibits ferromagnetic behavior with a Curie temperature above room temperature.[44] However, in our magnetization results shown in Fig. 3, there is no signature of ferromagnetism at room temperature. The magnetization data clearly indicates a paramagnetic nature, ruling out any significant contribution from an impurity phase to the magnetization, which might be due to negligible secondary phase. The inset in Fig. 3(b) shows the magnetization data at 4 K for low fields for better clarity. It can be noted that CoRuTiSn shows narrow hysteresis, indicating the soft ferromagnetic nature of the material.

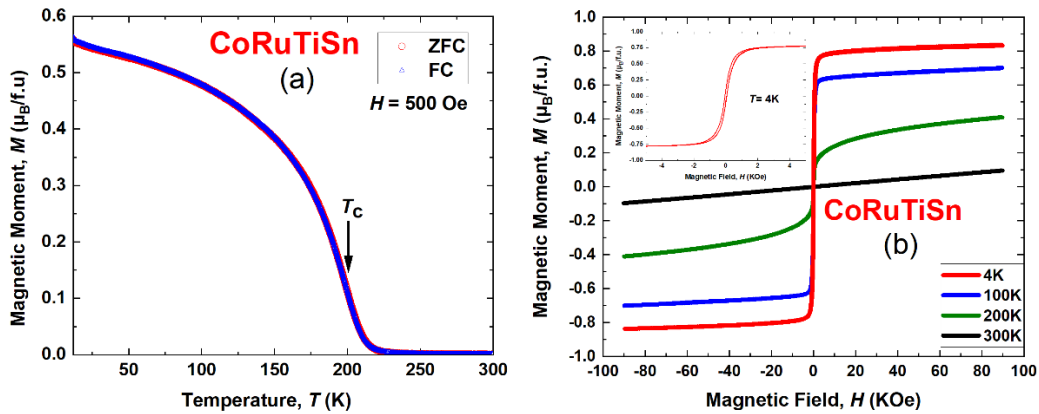

**Fig. 3.** (a) Temperature, $T$ dependence of magnetization, $M$ in zero-field cooled (ZFC) and field-cooled (FC) configurations at applied magnetic field, $H$ of 500 Oe. (b) The field dependence of $M$ as a function temperature for CoRuTiSn. The inset shows the hysteresis curve at 4 K.

### 3.3 Transport properties

In order to understand the electronic properties of CoRuTiSn, electrical transport measurements were carried out as a function of temperature and fields. Fig. 4 shows the temperature dependence of electrical resistivity, $\rho$ and electrical conductivity, $\sigma$ as a function of magnetic fields. It can be noted from Fig. 4(a) that the resistivity decreases with temperature, showing negative temperature coefficient, which confirms the semiconducting nature of CoRuTiSn. The application of field on resistivity does not show any significant effect. It has been reported that usually spin gapless semiconductors (SGS) show linear temperature dependence unlike conventional semiconductors.[7,16] Moreover, for SGS, a very weak temperature dependence for resistivity was observed. Careful observation of temperature dependence of resistivity/conductivity in Fig. 4 makes it clear that CoRuTiSn shows a strong temperature dependence, and it varies exponentially with temperature, which rules out possibility of SGS nature in CoRuTiSn. The temperature dependence of conductivity was fitted by equation (1) and is shown in inset of Fig. 4(b). [45-46]

$$\sigma(T) = \sigma_0 + \sigma_a e^{-E_g/K_B T} \qquad (1)$$

Here $K_B$ is Boltzmann constant, $E_g$ is bandgap, $\sigma_0$ is residual conductivity due to non-stoichiometry/defects at low temperatures and $\sigma_a$ is a zero-temperature conductivity for the semiconductor. The slight deviation between the experimental data and the theoretical fit could arise because the fitting equation (1) does not account for the spin polarization of the carriers. However, since the system is magnetic, spin-polarized carriers are expected, which is also supported by the band structure calculations. Moreover, a minor $Co_3Sn_2$ impurity may introduce defect states or secondary conduction channels, which could further contribute to the deviation from the theoretical fit.[47] A similar behavior was reported in other magnetic semiconductors such as $Mn_2CoAl$.[46] Bandgap determined from the fit of equation (1) to the experimental conductivity data was found to be ~ 60.4 meV. Slightly higher conductivity value and lower bandgap suggest that CoRuTiSn is a narrow bandgap semiconductor. The bandgap in CoRuTiSn is comparable to previously reported magnetic semiconductors such as CrVTiAl (63.4 meV), $Ti_{0.341}Zr_{0.343}Hf_{0.316}Co_{0.957}Pd_{0.043}Sb$ (70 meV), DyPdBi (50 meV) and $Mn_2CoAl$ (90 meV).[16, 45,46,48] Resistivity of $Co_3Sn_2$ was reported by K. Väyrynen *et al.* for different thickness and was in the range of µΩ-cm.[49] The resistivity of CoRuTiSn is in mΩ-cm range [Fig. 4(a)], which is significantly higher than that of $Co_3Sn_2$ (µΩ-cm range).[49] This indicates that the semiconducting transport properties are dominated by the CoRuTiSn matrix, with negligible contribution from the metallic $Co_3Sn_2$ phase.

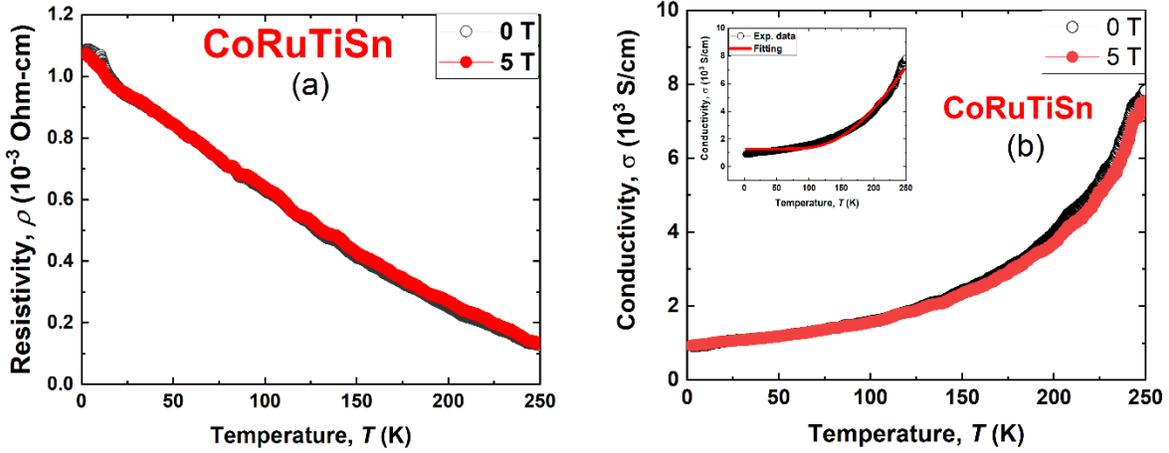

**Fig. 4.** (a) Temperature, $T$ dependence of (a) electrical resistivity, $\rho$ and (b) electrical conductivity, $\sigma$ at zero and 5T fields for CoRuTiSn. The Inset in (b) shows the fitting of conductivity data by equation (1).

Hall measurements for CoRuTiSn were carried out at 4 and 100 K by applying a magnetic field up to 2 T, perpendicular to the sample plane. The field dependence of Hall resistivity as a function of temperature is shown in Fig. 5(a). The squarer shape indicates anomalous Hall effect attributed to the ferromagnetic nature of the material, which resembles magnetic hysteresis loop. The total Hall resistivity, $\rho_{xy}$ which sums the resistivity due to ordinary Hall and anomalous Hall effect can be described as:[50]

$$\rho_{xy} = R_o \mu_0 H + R_s M \qquad (2)$$

where $R_o$ and $R_s$ are ordinary and anomalous Hall coefficients, respectively and $\mu_0$ is permeability in the free space. To calculate $R_o$, the Hall resistivity data was fitted in higher field regime at 4 K. The carrier concentration, $n$ was determined using the relation, $n = 1/qR_o$ (where $q$ is electronic charge). The charge carrier concentration of electrons and mobility, $\mu$ at 4 K in CoRuTiSn were found to be $10^{20}$ cm$^{-3}$ and $\sim 6.4$ cm$^2$/Vs, respectively, which is comparable with the values previously reported in various magnetic semiconductors such as CrVTiAl[16], CoFeCrGa [51] and Mn$_2$CoAl [46]. The mobility of electrons in CoRuTiSn is lower than observed in conventional semiconductors, which might be attributed to the presence of impurity states at the Fermi level.[16]

Fig. 5 (b) shows the anomalous Hall conductivity at 4 K temperature which was found to be ~ 4 S/cm and is comparable with the value reported in Mn$_2$CoAl thin film.[46]

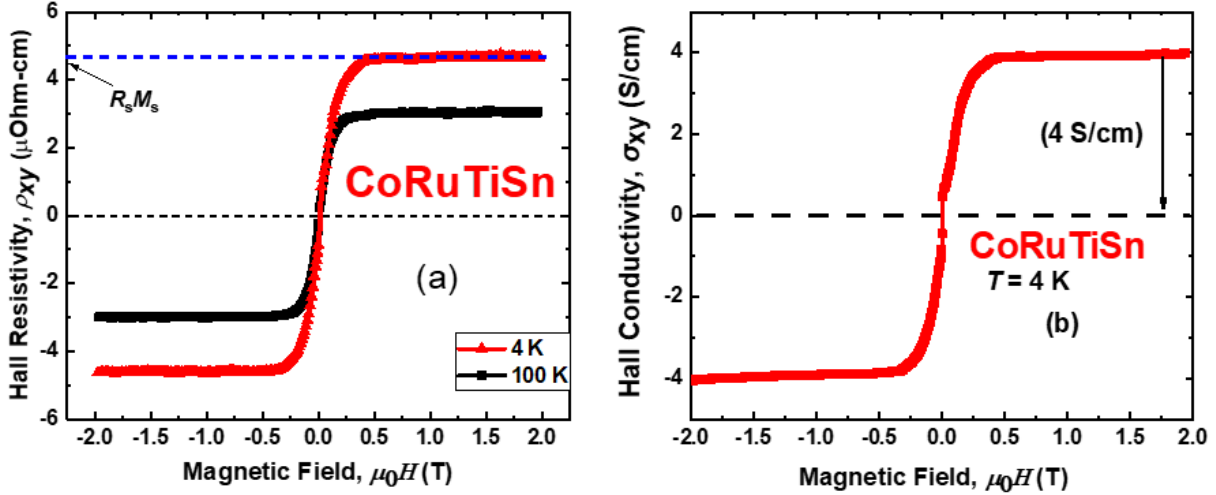

**Fig. 5.** The field dependence of Hall resistivity, $\rho_{xy}$ at 4 and 100 K. The dashed blue line is a guide for intercept at $H = 0$. (b) The field dependence of Hall conductivity, $\sigma_{xy}$ at 4 K.

The magnetoresistance (MR) is the change in the electrical resistance on the application of the magnetic field and is defined as[52]

$$\text{MR \%} = \frac{R(T,0)-R(T,H)}{R(T,0)} \times 100$$

here $R$ is electrical resistance. The field dependence of MR is shown in Fig. 6 at 4 K. It can be noted that the MR decreases with magnetic field due to suppression of spin fluctuation. The magnitude of MR is smaller and shows almost linear dependence with magnetic field. Similar results were reported in several other quaternary Heusler alloys.[53–55]

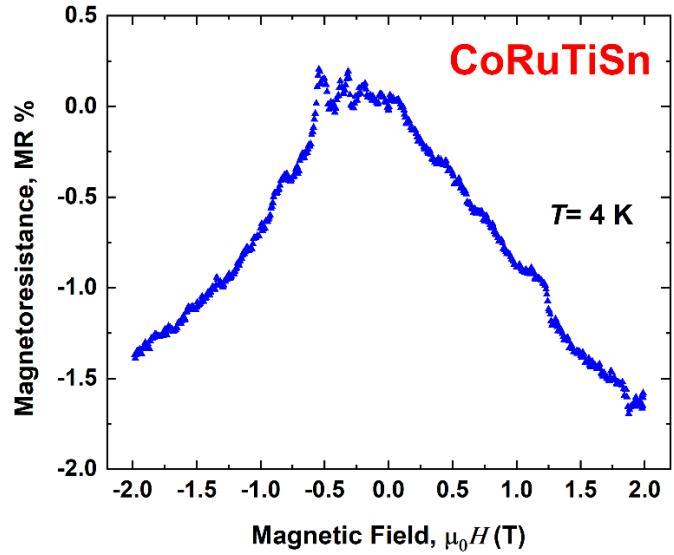

**Fig. 6.** The field dependence of magnetoresistance, MR at 4 K for CoRuTiSn.

### 3.4 Theoretical results

The phase stability of CoRuTiSn has been discussed in terms of the formation energy ($E_{form}$) and phase separation energy ($\Delta E$). $E_{form}$ signifies the stability of an alloy/compound against disintegrations into equilibrium phases of its constituent elements. On the other hand, $\Delta E$ stands for stability against decomposition into other alloys/compounds. For a stable compound it is expected both $E_{form}$ and $\Delta E$ are negative. The $E_{form}$ of CoRuTiSn is -0.46 eV/atom. The $\Delta E$ is marginally positive +0.01 eV/atom, considering the secondary phases such as $Co_2TiSn$, $Ru_2TiSn$. $Co_3Sn_2$, and RuTi. In a recent study Gao *et al.* reported, quaternary Heusler alloys could be synthesized experimentally if phase separation energy is less than 0.1 eV/atom.[56] Indeed CoFeCrAl was synthesized by different experimental groups despite positive phase separation energy.[57-58]

Ordered quaternary Heusler alloys have the *Y*-type crystal structure with a space group $F\bar{4}3m$. There are four interpenetrating fcc sublattices, 4a, 4b, 4c, and 4d. In the ordered structure Co, Ru, Ti and Sn atoms occupy 4a, 4b, 4c, and 4d sublattices, respectively. The optimized lattice parameter for the ordered structure is 6.214 Å, which is in reasonable agreement with the experimental value ($a = b = 6.21$ Å, and $c = 6.14$ Å). We calculated the total magnetic moment, which is 1.00 $\mu_B$, and consistent with the experimental value of 0.84 $\mu_B$. A slight mismatch with the experimental value can be accounted for in the secondary phase formation of $Co_3Sn_2$, thus the CoRuTiSn may deviate from stoichiometric composition slightly, which has not been considered in the calculations. It is to note that mainly Co atom contributes to the total magnetic moment, which is 0.91 $\mu_B$, whereas Ru, Ti and Sn carry tiny moments of 0.21 $\mu_B$, -0.05 $\mu_B$, and -0.02 $\mu_B$, respectively.

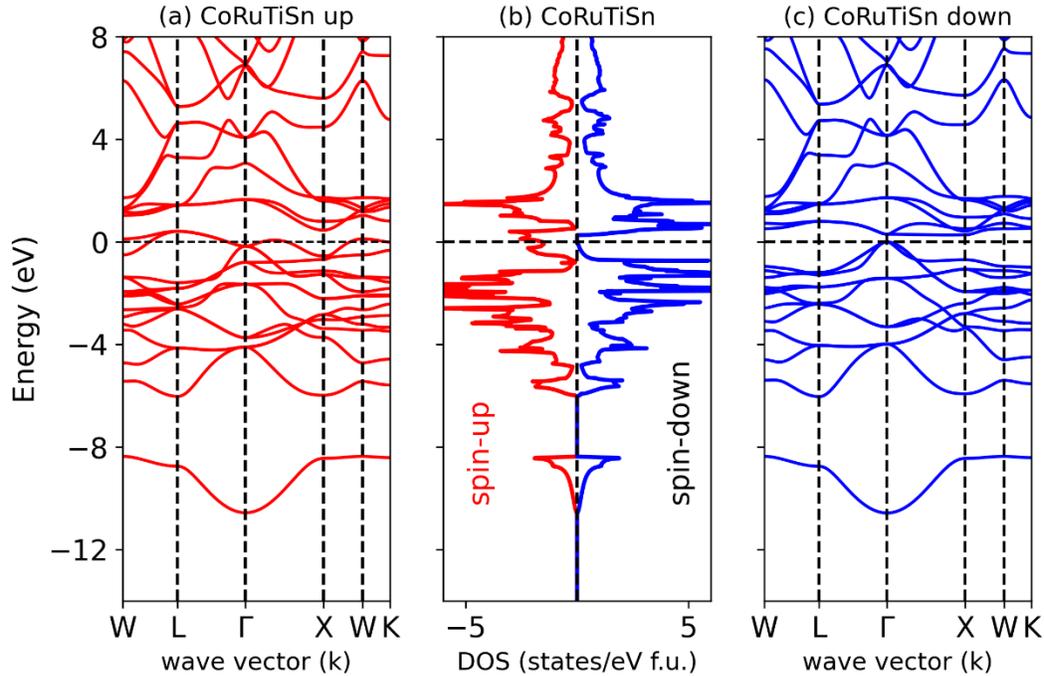

**Fig. 7.** Band dispersion curves and density-of-states (DOS) of ordered CoRuTiSn.

Fig. 7 shows the density-of-states (DOS) and band dispersion of CoRuTiSn along various high symmetry directions of the primitive Brillouin zone. It can be noted that the top of the valence band just touches the Fermi level leading to a pseudo gap at the Fermi level for the majority spin DOS. On the other hand, the minority spin channel is insulating. Thus, electronic structure resembles spin-gapless semiconductor but not an ideal one owing to the minute DOS at the majority spin channel at the Fermi level. The experimental bandgap of ~60.4 meV, determined from the temperature dependence of conductivity (Eq. (1)), represents an effective band gap, irrespective of spin channels. The theoretical prediction of spin gapless semiconducting nature (Fig. 7) assumes a cubic *Y*-type structure which is perfectly ordered. On the contrary, the sample is associated with a small tetragonal deformation and also may contain a minor impurity defects, which could modify the band structure, leading to a narrow bandgap magnetic semiconductor.[59]

Fig. 8 depicts the impact of various swapping disorders between constituent atoms of CoRuTiSn in terms of total energy and magnetic moment. We considered four different types of disorders, which include swapping of (i) Co and Ru atoms, (ii) Ti and Sn atoms, (iii) Co and Ru atoms, and Ti and Sn atoms, simultaneously, (iv) Ru and Ti atoms, leading to different types of crystal structures $L2_1$-I, $L2_1$-II, $B2$ and $XA$, respectively. Note that the perfectly ordered structure is $Y$-type. Fig. 8 compares the total energy for the disordered structures with respect to the ordered $Y$ structure (0 meV/f.u.). It is to note that the Co-Ru disordered structure ($L2_1$-I) is the energetically most favored (-88 meV/f.u.), followed by $Y$ (0 meV/f.u.), $B2$ (394 meV/f.u.), $L2_1$-II (445 meV/f.u.), $XA$ (1395 meV/f.u.). Total magnetic moment is almost robust against most of the disorders except the $XA$-type structure with Ru-Ti disorder, which is energetically very expensive, thus unlikely to realize in experiment.

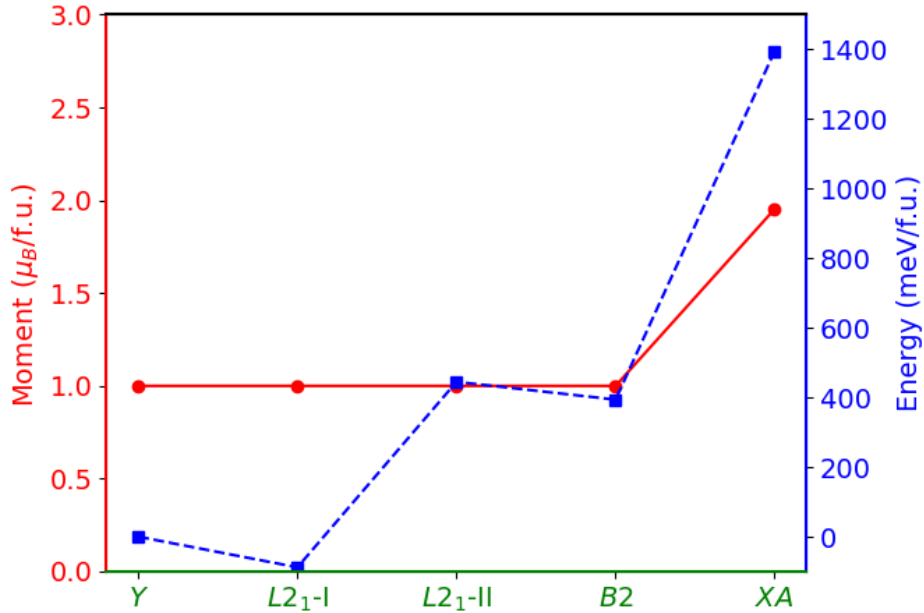

**Fig. 8.** Total magnetic moment (red circles) and total energy of ordered and disordered structures of CoRuTiSn.

Fig. 9 shows the DOS of the various disordered phases (solid red line), whereas the DOS of the ordered $Y$ structure (blue dotted line) is also presented for the sake of comparison. It is noteworthy that the half-metallic gap for the minority spin channel at the Fermi level remains robust in nature against the swapping disorders except for the case of swapping of Ru and Ti atoms.

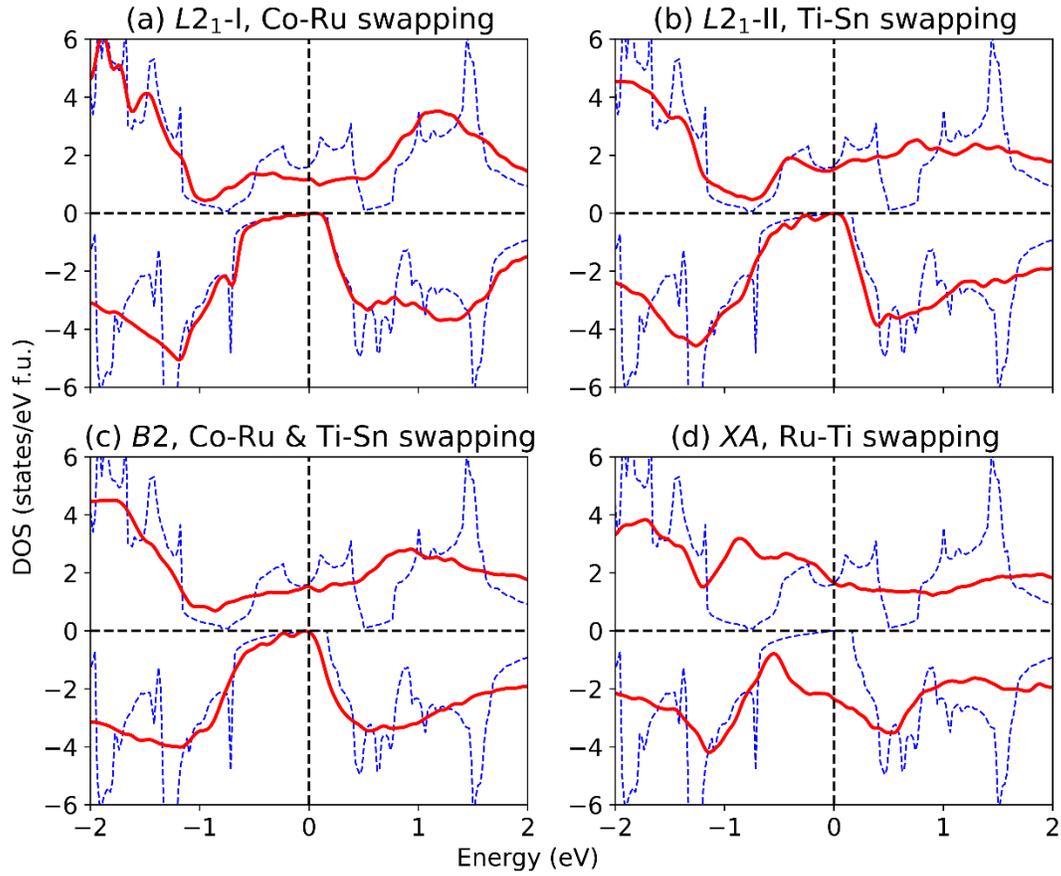

**Fig. 9.** Density-of-states (DOS) of the disordered structures (solid red line) compared with ordered CoRuTiSn (broken blue lines).

## 4. Conclusions

We synthesized CoRuTiSn by an arc melt technique and studied its structural, magnetic and electronic properties along with theoretical calculations. The XRD pattern reveals tetragonal crystal structure. Magnetic measurements show low moment soft ferromagnetism in CoRuTiSn with a Curie temperature of about 200 K. The temperature dependence of electrical resistivity indicates semiconducting behavior with a band gap of ~ 60.4 meV. The carrier concentration and mobility determined from Hall measurements data was found to be comparable with other magnetic semiconductors. Magnetoresistance shows almost linear dependence on field at 4 K. Observation of semiconducting as well as low moment properties in CoFeRuSn could make this material promising for semiconducting spintronic applications.

**Acknowledgments:**

The authors express their gratitude to Prof. K. G. Suresh, IIT Bombay, for experimental support in material synthesis and fruitful discussion. S.G. acknowledges the financial support from Anusandhan National Research Foundation (ANRF), New Delhi under the project SUR/2022/004713 for carrying out this research work. This work was supported in part by X-NICS (Grant No. JPJ011438) from MEXT and by Center for Science and Innovation in Spintronics (CSIS), Tohoku University.